\documentclass[runningheads]{llncs}

\title{Online Algorithms for Self Organizing Sequential Search - A Survey}
\date{}
\titlerunning{Literature Survey}
\author{Rakesh Mohanty and N. S. Narayanaswamy }
\authorrunning{Rakesh Mohanty and N. S. Narayanaswamy }
\institute{Department of Computer Science and Engineering, Indian Institute of Technology, Chennai-600036, India. \\
\email{\{rakesh.iitmphd, narayanaswamy @ gmail.com \}}}
\begin{document}
\maketitle
\begin{abstract} The main objective of this survey is to present the important theoretical and experimental results contributed till date in the area of online algorithms for the self organizing sequential search problem, also popularly known as the List Update Problem(LUP) in a chronological way. The survey includes competitiveness results of deterministic and randomized online algorithms and complexity results of optimal off line algorithms for the list update problem. We also present the results associated with list update with look ahead, list update with locality of reference and other variants of the list update problem.  We investigate research issues, explore scope of future work associated with each issue so that future researchers can find it useful to work on.
\end{abstract}
\section{Introduction}
Design of  efficient algorithms under the limitations of accessibility of data is a real challenging research area in computer science.  The traditional design and analysis of algorithms assumes that an algorithm generates output with complete knowledge of the entire input.  However, this assumption is often unrealistic in practical applications.  Many of the algorithmic problems that arise in real life are \emph{online} in the sense that the input is only partially available and some relevant input data arrives in the future being not accessible at present.  Online problems arise in many areas of computer science such as resource management in operating systems, routing in communication networks, paging in virtual memory, scheduling and data structures\cite{Alb2003}.  An \emph{online algorithm} must generate output without knowledge of the entire input.  The quality of an online algorithm is usually evaluated using an approach called \emph{competitive analysis}.  In this approach, an online algorithm for a problem is compared to an optimum off-line algorithm.  An \emph {optimum off-line algorithm} knows the entire input sequence in advance and can process it with minimum cost.  Given an input sequence $\sigma$, let ALG($\sigma$) denote the cost incurred by online algorithm ALG in processing $\sigma$ and OPT($\sigma$) denote the cost incurred by optimal off-line algorithm  OPT in processing $\sigma$.  Then the algorithm ALG is \emph {c-competitive}, if there exists a nonnegative constant $\alpha$ such that ALG($\sigma$) $\leq$ c.OPT($\sigma$) +  $\alpha$  for all input sequences $\sigma$.  Here c is called the \emph {Competitive Ratio(CR)} of the online algorithm ALG.  The competitive ratio is always at least 1 and the smaller it is, the better ALG performs with respect to OPT. In computer science, {\em sequential search}  is a search algorithm for searching a list of unsorted data items for a particular value and is one of the simplest basic search method. In sequential search, we start from the beginning of the list, scan each item one by one down the list, until the desired item is found or we have reached the end of the list. A {\em self organizing sequential search algorithm} may rearrange the order of the items in the list in some fashion just after an item is searched and place the more frequently searched item closer to the front of the list to reduce future search times. The {\em self organizing sequential search problem}, popularly known as the list update problem is of significant practical interest in the context of self-organizing data structures.{\em Self organizing data structures} reorganize their structure while processing a sequence of operations.  The purpose of this reorganization is to guarantee the efficiency of future operations. Self-organizing rules are effective because they take advantage of locality of reference found in real time systems.\\
{\em The list update problem} is concerned with maintaining a dictionary as an unsorted linear list. The {\em dictionary} is an abstract data type, which is frequently used in practice because of its great simplicity. An {\em abstract data type} is a set of data values and associated operations that are precisely specified independent of any particular implementation. When the dictionary is small (such as e.g. organizing the list of identifiers maintained by a compiler or organizing collisions in a hash table or when there is no space to implement time efficient space consuming data structures), linear list is the implementation of choice.  An unsorted linear list is one of the simplest data structure on which we can perform insertion, deletion and access operations.  To perform access, the list has to be traversed linearly from the start of the list until the requested item is found. The insertion operation is performed by adding an item immediately after the last item of the list and deletion of an item is performed by simply removing the item from the list after accessing the item.\\
{\em Problem Statement :}  Given an unsorted linear list L of  l distinct items and a finite sequence of input requests $\sigma$ of size n, where each request is either an access, insert or delete operation on an item in the list.  As insert and delete operation can be regarded as a special case of access, we can only perform access operation and refer the list update problem as the {\em list accessing problem}.  A request is said to be serviced, if we perform a search or access operation of the requested item in the list by incurring some access cost. After accessing an item, we rearrange the items in the list by incurring some reorganization cost before the next access. Our goal is to minimize the total reorganization and access cost while  serving a given request sequence $\sigma$ on a list of size l.\\
{\em Motivation and Applications :} List update techniques have been extensively used in practice when storing and maintaining small dictionaries.  They are also used to develop fast and simple algorithms for computing point maxima and convex hulls in computational geometry.  Another important application of list update is for data compression\cite{BE1998}.\\
{\em Historical overview of research work :}  The list update problem is of significant theoretical and practical interest for the last four decades.  As per our knowledge, study of list update techniques was initiated by the pioneering work of McCabe \cite{McCabe1965} in 1965.  He investigated the problem of maintaining a sequential file and developed two algorithms Move-To-Front (MTF) and Transpose.  From 1965 to 1985, the list update problem was studied by many researchers \cite{ANW1982,BM1985,Bit1979,CHS1985,Fre1984,GMS1979,GMS1981,Riv1976,Ten1978,TN1982} under the assumption that a request sequence is generated by a probability distribution.  Hester and Hirschberg\cite{HH1985} have provided an extensive survey of average case analysis such list update algorithms.  The seminal paper by Sleator and Tarjan\cite{ST1985} in 1985 made the competitive analysis of online algorithms very popular.  Quite a few of the competitive analysis of the list update problem were carried out without any specific knowledge of the nature of the optimal off-line algorithm. The first use of randomization and the demonstration of its advantage in the competitive analysis context was done by Borodin, Linial and Saks\cite{BLS1985} with respect to metrical task systems in 1987.  The first study of randomized online algorithms for the list update problem was initiated by Irani, Reingold, Westbrook and Sleator\cite{IRSW1991} in 1990. Teia\cite{Tei1993} proved a lower bound of 1.5 on the competitive ratio of any randomized online algorithm in 1993. The best randomized online algorithms with competitive ratio 1.6 by COMB is due to Albers and et. al \cite{AVW1995}.A detailed study of off-line algorithms for the list update problem was published  by Westbrook and Reingold\cite{RW1996} in 1996. Ambuhl and et. al.\cite{AGV2001} proved a more stronger lower bound of 1.50084 for the partial cost model.  Ambuhl\cite{Amb2000} also proved that off-line list update is NP-hard in 2000 by showing a reduction from the Minimum Feedback Arc Set Problem. Martinez and Roura\cite{MR2000} in 2000 proved that, under a more realistic model no online algorithm including MTF can be c-competitive for any constant c. Bachrach and et. al. have provided an extensive theoretical and experimental study of online list accessing algorithms in 2002\cite{BER2002}.\\
{\em Contributions and Organization of this survey :} The aim of this survey is to present the theoretical and experimental research work that has been done in the area of online algorithms for the {\em self organizing sequential search problem} also known as the {\em list update problem}. In section 2, the standard list update cost model, both deterministic and randomized online algorithms  and concept of adversary is introduced. In section 3 and 4 , we present the well known deterministic and randomized online algorithms for the list update problem with competitiveness results.  Section 5 includes online algorithms for list update with look ahead and section 6 includes online algorithms for list update with locality of reference. Various empirical studies are highlighted in section 7 and variants of the list update problem are mentioned in section 8. Off-line algorithms for list update is covered in section 9. We explore the research directions and investigate the scope of future work associated each research issue of the list update problem in each of the above sections 3-9. Finally section 10 presents a concluding remark.
\section{Preliminaries}
{\em The list update cost model :} This model is a standard cost model defined by Sleator and Tarjan.  A list L consists of l labelled unsorted items.  Operations to be performed are access, insert and  delete.  Accessing the item in the list which is in $i^{th}$ position from the front of the list costs i.  If the item is not present in the list, then the access cost is l.  Cost of servicing each request for item x is 1 + (number of items before x in the list).  If x is not present in the list, service cost for each request is l + 1.  Inserting an item costs l + 1.  Deleting the  $i^{th}$ item costs  i. \emph {Free transposition} : After item x is accessed or inserted, it is moved to any location towards the front of the list with no cost.  \emph {Paid transposition} is the exchange in position of adjacent items in the list at a cost of 1.  The goal is to reorganize  the list L by performing free and paid transpositions that minimize access and reorganization cost.  An algorithm can reorganize the list at any time.  Reorganization cost is measured by minimum number of transpositions of consecutive items in the list.  \emph {Static list update model} is one in which the number of items is fixed in the list and only access operation is performed.  \emph {Dynamic list update model} is one in which the number of items in the list vary and all the three operations such as access, insert and delete can be performed. Until unless specified, in this paper we consider the static list update and standard cost model.\\
{\em List update algorithm :} The online list update algorithms can be classified in to two types such as deterministic and randomized. A {\em deterministic online algorithms} updates the list based on current and past requests. An \emph {adversary}, often called the \emph {off-line player}, based on the knowledge of the algorithm used by online algorithm tries to make the task costly to the on-line algorithm  by constructing the worst possible input.  An \emph {oblivious adversary} must construct the request sequence in advance and pays optimally to service the request sequence.  An \emph {adaptive online adversary} serves the current request on-line and then chooses the next request based on online algorithms action so far.  An \emph {adaptive off-line adversary} chooses the next request based on online algorithm's action thus far, but pays optimal off-line cost to service the resulting request sequence.  A \emph {randomized strategy} is defined by a probability distribution over the set of all deterministic strategies.  In \emph {online randomized algorithm}, before serving the first request, one of the deterministic strategies is chosen according to probability distribution.  Let ALG be the randomized algorithm which is online.  Based on the probability distribution ALG uses, the oblivious adversary must choose a finite request sequence $\sigma$ in advance.  ALG is \emph {c-competitive against an oblivious adversary}, if for every such $\sigma$, \emph {E[ALG($\sigma$)] $\leq$  c.OPT($\sigma$)  + $\alpha$}, where $\alpha$ is a constant independent of $\sigma$.  Here OPT is not a random variable, hence off-line player does not have info about the outcomes of random choices made by online algorithm.  A randomized on-line algorithm ALG is \emph {c-competitive against an adaptive online adversary}, if there is a constant $\alpha$  such that for all lists and all adversaries, A', \emph{ E[A($\sigma$) - c. A'($\sigma$)] $ \leq$  $\alpha$}.
\section{Deterministic online algorithms}
In this section we present some well known deterministic online algorithms for the list update problem as proposed in the literature. A summary of important results is presented in Table 1.
\begin{itemize}
\item {\em Move-To-Front (MTF) :} After accessing an item x in the list, move it to the front of the list without changing the order of other items in the list.
\item {\em Transpose (TRANS) :} Transpose the accessed item x with the immediately preceding item in the list. 
\item {\em Frequency Count (FC) :} Initialize frequency counter of each item to 0. After accessing an item, increment its frequency counter by one. Rearrange the items in non-increasing order of their frequencies.
\item {\em MTF2 :} Move x to the front of the list, if the position of accessed item x is even. 
\item {\em MHD(k): } Move accessed item x forward k positions in the list.
\item {\em MF(k) : } Move accessed item x from its current $i^{th}$ position to $\lceil$ (i/k) $\rceil$ - 1 positions towards the front of the list.
\item {\em Time Stamp (TS) :} Insert accessed item x in front of the first item y that precedes x on the list and was requested at most once since the last request for x. Do nothing, if there is no such y, or if x is requested for the first time.
\item {\em Pass-Recent-Item(m) or PRI(m) :} Move accessed item x forward just in front of item z on the list that was requested at most m times since the last request for x. Do nothing, if there is no such z, or if x is requested for the first time in the list.
\item {\em Move-to-Recent-Item(m) or MRI(m) :} Move accessed item x forward just after the last item z  on the list that is in front of x and has been requested at least m+1 times since the last request for x. If there is no such z, move x to the front or if x has been requested for the first time, do nothing.
\end{itemize}
\begin{center}
  \begin{tabular}{ | c | c | c | }  \hline
     {\bf Algorithm} &  {\bf Competitiveness Result} & {\bf Researcher(s) and Year}  \\ \hline
      MTF &  2-competitive & Sleator, Tarjan \cite{ST1985} [1985]\\ \hline
      MTF  &   2- 2/(l+1): Lower Bound & Karp, Raghavan\\ \hline
      MTF  &   2- 2/(l+1): Upper Bound & Irani\cite{Ir1991}  [1991]\\ \hline
      TRANS & 2l/3 : Lower Bound & Sleator, Tarjan  \cite{ST1985} [1985]\\    \hline
      FC  &  (l+1)/2 : Lower Bound & Sleator, Tarjan  \cite{ST1985} [1985]\\ \hline
      MTF2 &  2-competitive & Bachrach, El-Yaniv\cite{BE1997} [1997]\\ \hline
      MF(k) & 2k-competitive& Sleator, Tarjan \cite{ST1985} [1985]\\ \hline
      TS(0) & 2-competitive & Albers\cite{Alb1995}  [1995]\\ \hline
      MRI(k)& 2-competitive & El-Yaniv\cite{EY1996} [1996]\\ \hline 
      PRI(m) & 3-competitive, for m$\geq$ 1 & Bachrach et. al. \cite{BER2002} [2002]\\ \hline
  \end{tabular}
\end{center}
\begin{center}{\bf Table 1 : Competitiveness of Deterministic Online algorithms}\end{center}
{\bf Research Issues  : }
\begin{itemize}
\item Define the List update cost model more rigorously to achieve better analysis results.
\item Generation, classification and characterization of different types of request sequences that can model real world inputs.
\item Determination of optimal deterministic competitive ratio for dynamic list update problem.
\item Design of deterministic online algorithms that affect locality of reference for real life inputs.
\end{itemize}
\section{Randomized Online Algorithms}
A number of randomized online algorithms have already been proposed by researchers for the list update problem. We present some well known algorithms in the following section and a summary of results in Table 2.
\begin{itemize} 
 \item {\em SPLIT : } For each item x, maintain a pointer p(x) pointing to some item in the list. With probability 1/2, move accessed item x to the front and with probability 1/2, insert x in front of p(x). Then set p(x) to the first item in the list.
\item {\em BIT :} For each item x in the list, maintain a mod-2 counter b(x). Initially set b(x) to 0 or 1 randomly, independently, uniformly. Then complement b(x). If complement of b(x) = 1, move accessed item x to the front of the list.
\item {\em RMTF :} With probability 1/2, move the accessed item x to the front.
\item {\em COUNTER(s, S) :} For each item x,  maintain a mod-s counter c(x)$\in$ $\{$0, 1, .., s-1$\}$. To access item x, decrement c(x) by 1(mod s). If c(x) $\in$ S, move x to front.
\item {\em RST(s,D) :} For each x,  maintain a mod-s counter c(x) $\in$ $\{$0, 1, ..., s-1$\}$ set randomly with probability D(i). For accessing item x, decrement c(x) by 1(mod s). Then if c(x) = 0, move x to the front and randomly reset c(x) using D.
\item {\em TS(p) :} For accessing item x, with probability p execute (i) move x to the front; and with probability 1-p execute (ii).(ii) let y be the first item in the list such that either (a) y was not requested since the last request for x or (b) y was requested exactly once since the last request for x and that request for y was served by the algorithm using step (ii). Insert x just in front of y. Do nothing, if there is no such y or this is the first request to x.\cite{Alb1995}.
\item {\em COMB :} Before serving any request, choose algorithm  BIT with probability 4/5 and algorithm TS with Probability 1/5. Serve the entire request sequence with the chosen algorithms.
\end{itemize}
{\bf Research Issues :}
\begin{itemize}\item Close or diminish the gap of [1.5, 1.6] in competitive ratio for the randomized online algorithms against oblivious adversary.
\item Determine the competitive ratio of $RMTF_p$ for each p $\in$ (0,1).
\item Close or diminish the gap of [1.625, 1.75] for the BIT algorithm.
\end{itemize}
\begin{center}
\begin{tabular}{ | c | c | c |  } \hline
   {\bf Algorithm} &   {\bf Result} & {\bf Researcher(s)} \\ \hline
      -       & Lower Bound: 1.18 & Karp, Raghavan [1990]\\ \hline
    SPLIT  &  CR : 1.875 & Irani\cite{Ir1991}  [1991] \\ \hline
    -       &  1.5 : Lower Bound & Teia\cite{Tei1993}  [1993] \\ \hline
    BIT & CR : 1.75  & Reingold, Westbrook, Sleator\cite{RWS1994}[1994]\\    \hline
    RST(s, D) & CR : 1.732 & -  \\ \hline
    TS(p) & 1.62-competitive & Albers\cite{Alb1995}  [1995]\\ \hline
    COMB & CR : 1.6 : Best Upper bound & Albers, Von stengel, Werchner\cite{AVW1995}  [1995]\\ \hline
    COUNTER &  CR : 85/49 & Albers, Mitzenmacher\cite{AM1997}  [1997]\\ \hline
     RMTF  &  CR : $\geq$ 2 & Garefalakis\cite{Gar1997}  [1997]\\ \hline
       - & Stronger Lower Bound : 1.5 & Hagerup\cite{Hag2007} [2007] \\ \hline
\end{tabular}
\end{center}
\begin{center}{\bf Table 2 : Competitiveness of Randomized Online algorithms}\end{center}
\section{\bf List Update with Look Ahead}
In the list update problem with look ahead, the online algorithm knows some future requests in th e form of blocks of requests of variable size.  Albers[1998] studied the influence of look ahead in the list update  problem and introduced two different models of look ahead such as - weak look ahead and strong look ahead\cite{Alb1994}.\\
{\bf  Weak  look ahead: } In this model, the online algorithm knows the present request and next l future requests. When processing $\sigma$(t), online algorithm only knows $\sigma$(t+1), $\sigma$(t+2), ..., $\sigma$(t+l) but does not know $\sigma$(s) such that s $\geq$ t+l+1.\\
{\bf Strong look ahead :} In this model, the online algorithm knows the present request and a sequence of future requests which contains l pairwise distinct items which differ from the present requested item.
\begin{center}
\begin{tabular} { |c|c|c|} \hline
           Model    &  Lower Bound & Upper Bound\\  \hline
           Strong look ahead &   2-(l+2)/(n+1) &  2-(2/3)(l+2)/(2n-l) \\ \hline
           Weak Look ahead &  2- 2$\sqrt(4K^22K)$+ 4K &  $2- (2/3)\sqrt(K^2+2K)$ - K \\ \hline
\end{tabular}\end{center}
\begin{center} {\bf Table 3 : Summary of Competitiveness Results }\end{center}
{\bf Research Issues}
\begin{itemize} \item Develop some alternate model for look ahead other than weak and strong model for analysis.
\item Design of semi-online algorithm for varying size of the look ahead.
\item Design and analysis of randomized online algorithm for LUP with strong and weak look ahead.
\item Design of online algorithms that are competitive  against dynamic off-line algorithms.
\item Design of deterministic online algorithms for n-element list with k-lookahead problem.
\item Tighten the gap between upper and lower bounds in the weak and strong look ahead model.
\end{itemize}
\section{\bf List Update with Locality of Reference :}
A study of locality of reference in list update was initiated by Angelepoulos et al. \cite{ADL2008}, where they adapted a locality model introduced for the paging problem and proved that MTF is superior to other algorithms. A comprehensive study of list update with locality of reference was presented recently by Albers and Lauer\cite{AL2008}. Their theoretical and experimental work provided a refined analysis of the problem with a new model based on the concept of runs in which theoretical and empirical results match or nearly match.\\\\
{\bf Research Issues :}
\begin{itemize}
\item Theoretical and experimental study of online algorithms list update with locality of reference.
\item Developing a new cost model to capture the locality of reference in LUP.
\item Developing improved online algorithm for list update with locality of reference.
\end{itemize}
\section{\bf Empirical Studies}
A number of experimental studies have been conducted by various  researchers on the list update problem. Tannenbaum [1978] tested the performance of various algorithms from the MOVE-AHEAD(k) family with respect to request sequences generated by Zipf's law.  Bentley and McGeoch[1985] have tested the performance of MTF, FC, TRANS with respect to request sequences generated from several text files and found that FC is always superior to TRANS and MTF is often superior to FC. Albers and Mitzenmacher [1995] have compared the compression performance of TS and MTF with respect to Calgary Corpus by considering both word and byte parsing scheme.  Burrows and Wheeler[1994] tested the performance of an MTF compressor via block sorting transformation and Grinberg et al.[1995] tested the performance of MTF compressors that uses secondary lists.  Bachrach and El-Yaniv [1997] have conducted an extensive experimental study of  a large number of different online list accessing algorithms based on access cost performance and compression performance.  They considered a wide range of request sequences and used an experimental approach to examine the influence of locality of reference in request sequences on the performance of online list accessing algorithms.  They showed that the degree of locality has a considerable influence on algorithms' cost and  their ranking.  The two extreme cases studied so far concern with the the inputs as an independent observations of a probability distribution and the inputs generated by an adversary aiming to maximize the competitive ratio of the algorithm. Although a number of interesting theoretical results have been obtained in the analysis of list accessing problem, an experimental feedback could have lead to better and more realistic theory\cite{BE1997}.
\paragraph{} We present below a summary of research work done till date on experimental analysis of empirical studies of self orgainizning sequential search algorithms.
\begin{center}
\begin{tabular}{|c|c|}\hline
{\bf Researcher(s)} & {\bf Brief description of  Work} \\ \hline 
Tenenbaum[1978]\cite{Ten1978} & Tested performance of MHD(k) family \\
                              & algorithms with respect to request\\
                              & sequences distributed by Zipf's law. \\  \hline
Bentley, McGeoch[1985]\cite{BM1985}& Tested performance of MTF, FC, TRANS \\
                                   & with respect to request sequences \\
                                   & generated from several text files.\\ \hline
Burrows, Wheeler[1994] & Tested performance of MTF compressor \\
                                    & via block sorting transformation.\\ \hline
Grinberg, Rajagopalan et al.[1995]     & Tested performance of MTF compressors\\
                                                 & that uses secondary lists. \\ \hline
Albers, Mitzenmacher[1995]\cite{AM1997}& Compared compression performance of TS\\
                                       & with that of MTF for Calgary Corpus by\\
                                       & considering word and character parsing.\\ \hline
Bachrach, El-Yaniv[1997]\cite{BE1997} & Tested performance of deterministic and \\
                                      & randomized online list accessing algorithms\\
                                      &  with respect to dictionary maintenance and \\
                                      & compression applications.\\ \hline
Bachrach, El-Yaniv, Reinstadtler[2002]\cite{BER2002} & Extensive empirical study of a large set \\
                                     & of online list accessing algorithms including \\
                                     & MRI and PRI families\\ \hline
\end{tabular}
\end{center}
\begin{center} {\bf Table 4. A Summary of Experimental Analysis }\end{center}
{\bf Research Issues :} 
\begin{itemize}
 \item Determination of some new alternate metric for measuring the performance of list accessing algorithms.
\item Devise a meaningful, quantitative measure of locality of reference that could be used to classify request sequences.
\item Design an experimental set up which will cover a wide range of request sequences.
 \item  Investigate the correlation between various algorithms and their performance with respect to the request sequences for real life inputs. 
\item  Design an appropriate corpus for testing the performance of data structures and algorithms used for dictionary maintainence. 
\item  An Experimental study for dynamic transitions between different basic list accessing algorithms for adapting to changing levels of locality of reference.
\end{itemize}
\section{\bf List Update Variants}
\subsection{\bf Parallel List Update Problem}
Given a set S of n.m elements arranged in n lists $L_1$, $L_2$, ...., $L_n$ of m elements each, with m $>$ n and a sequence of requests $\sigma$ = $R_1$, $R_2$, ..., $R_n$, where each request $R_i$ $\subset$ S is a set of n elements.  Goal is to search the elements of each $R_i$ in parallel and then update the lists in order to process $\sigma$ efficiently.  Luccio and Pedrotti in 1994 \cite{LP1994} have proposed ($n^2$+1)-competitive algorithm against a static optimal algorithm with a lower bound of 2n.   They also developed a randomized algorithm with a competitive factor of (9/2)n against an oblivious adversary with a lower bound of (3/2)n. 
\subsection{\bf Distributed List Update Problem}
Here the list update problem is studied in a distributed environment where the set of items is partitioned across 2 processors and in which the cost of accessing an item  is a combination of a list searching cost and a communication cost.  Shende and Simha in 1995 \cite{SS1995} have studied the distributed list update problem under 2 models. In the Global Knowledge model, the algorithm knows the partition of items in advance, while in the Local Knowledge model, the algorithm has no knowledge regarding the partition of items.  They have proved that Distributed Move-To-Front(DMTF) is not competitive in Local Knowledge model and 4-competitive in case of Global Knowledge model.
\subsection{\bf Relaxed List Update Problem}
Relaxed List Update Problem(RLUP) is a variant of LUP, in which cost to access the jth item $x_j$ is $c_j$ where $c_i$ $\leq$ $c_{i+1}$ for all i.  After accessing, $x_j$ can be repeatedly swapped at no cost, with any item that precedes it in the list.  This problem was introduced by Agrawal et al.[1987] as  a model for management of hierarchical memory that consists of a number of caches of increasing size and access time.  Chrobak and Noga[1998] have developed an optimal off-line algorithm for RLUP and showed a characterization of work functions for RLUP.  They proved that MTF is optimally competitive for RLUP with any cost function and also provided a lower bound on the competitive ratio of online algorithms for RLUP.
\section{\bf Off-line List Update}
Off-line algorithms for the list  update problem are useful in the study of competitive online algorithms. We present below a summary of results for the off-line list update problem. 
\begin{center}
  \begin{tabular}{ | c | c | c | } \hline
       {\bf Complexity Result} & {\bf Researcher(s)}  & {\bf Year}\\ \hline
      $\theta$(n$(l!)^2$) & Manasse, McGeoch, Sleator\cite{MMS1988} & 1988\\ \hline
      $\theta$(n$2^l$(l-1)!) & Reingold, Westbrook\cite{RW1996} & 1996\\ \hline
      NP-Hard  &  Ambuhl\cite{Amb2000}  &  2000 \\  \hline 
      $\theta$(n$l^3$l!) & Pietzrak\cite{Pie2001} & 2001 \\ \hline 
      O($2^l$l!f(l)+m+r) & Hagerup\cite{Hag2007} & 2007 \\ \hline 
\end{tabular}
\end{center}
\begin{center}{\bf Table 5. Summary of Results of Off-line List Update}\end{center}
{\bf Research Issues :}
\begin{itemize}
\item Finding the best way to rearrange the list most effectively with optimal cost.
\item Derive a lower bound for computing an optimal solution.
\item Find an alternate reduction technique to prove NP-hardness of off-line list update.
\item Design of  efficient exact algorithms for list update. 
 \item Investigation of better optimum off-line algorithms for List Update Problem.
\item New characterization for optimal off-line algorithms. 
\item Design of improved approximation algorithms with a bounded approximation ratio.
\end{itemize}
\section{\bf Concluding Remarks}
We present a summary of best known bounds for both deterministic and randomized online algorithms  for list of different sizes for the list update problem till date in the following table.
\begin{center}
\begin{tabular}{|c|c|c|c|}\hline
list size & Lower Bound & Upper Bound & Remarks \\ \hline
2 & 1.125    &   1.125           & tight \\ \hline
3 & 1.2      &   1.2             & tight \\ \hline
4 & 1.25     &   1.33            & gap : 0.08 \\ \hline
5 & 1.25     &   1.6             & gap: 0.35 \\ \hline
6 & 1.28     &   1.6             & gap : 0.32 \\ \hline
- & -        &   -               & -        \\  \hline
l & 1.5-2/(l+3)& 1.6             & gap : 0.1 \\ \hline
\end{tabular}
\end{center}
\begin{center}{ \bf Table 6.  Best Known Bounds for Online List Update} \end{center}
The goal of future research must be to find improved online algorithms for the list update problem by designing  some new principle that will lead to better results and tighten the currently existing lower and upper bounds 
\bibliographystyle{splncs}
\bibliography{lit_survey}
\end{document}